\documentclass[a4paper,11pt]{article}

\usepackage[utf8]{inputenc}
\usepackage[T1]{fontenc}
\usepackage{lmodern}
\usepackage{geometry}
\usepackage{graphicx}
\usepackage{authblk}
\usepackage{url}
\usepackage{cite}
\usepackage{fancyhdr}
\usepackage[hidelinks]{hyperref}
\usepackage{orcidlink}
\usepackage{subcaption} 
\usepackage{amsmath} 
\usepackage{slashed} 
\usepackage{hyperref} 

\fancypagestyle{firstpage}{
	\fancyhf{}
	\fancyhead[C]{\includegraphics[height=2.2cm]{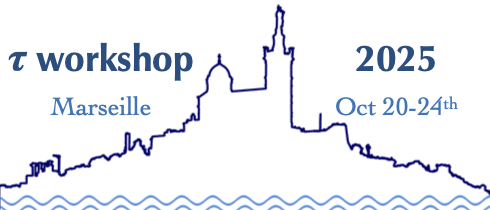}\\ \textit{Proceedings of the 18th International Workshop on Tau 
Lepton Physics}
	} 

}

\title{Simulation of tau decays, ambiguities and anomalous couplings}
\author[1]{Zbigniew Was\,\orcidlink{0000-0002-1615-9038}\thanks{Presenting author.}, Ananya 
Tapadar\,\orcidlink{0009-0007-3514-1427}, J. M. John\,\orcidlink{0000-0001-6879-7464}}
\author[2]{S. Banerjee\,\orcidlink{0000-0001-8852-2409}}

\affil[1]{Institute of Nuclear Physics, Polish Academy of Sciences, Cracow, Poland}
\affil[2]{University of Louisville, Louisville, USA}

\date{December 21, 2025}

\begin{document}

\maketitle
\thispagestyle{firstpage}
\vspace{-1cm}
\begin{center}
	\textit{IFJPAN-IV-2025-25}
\end{center}
\vspace{0.5cm}

\begin{abstract}
 From the perspective of low energy $\tau$ decays and radiative corrections in decays, not much has changed since the last, {\tt 
Tau23} conference. Also, {\tt TAUOLA}, $\tau$ decay library, and {\tt PHOTOS} for radiative corrections in decays have not changed 
 much, neither for QED nor for New Physics processes application. Progress was
 in the domain of flexibilities for applications and in the
domain of { \it New Physicists.}  There was progress with  use of   exclusive exponentiation ({\tt KKMC Monte Carlo}) to evaluate 
quality of $\tau$ pair formation and  decay factorization, with  Collins-Soper and  {\tt Mustraal} frames. Ambiguities  for  New Physics 
or other processes could be thus addressed. Associated projects of  {\tt KKMC} f77 and {\tt KKMCee}  C++, are presented:  
  (i) $\pi^+\pi^-$ pair production at {\tt Belle II}  energies, interesting
  input for $\Pi_{\gamma\gamma}(s)$,
  (ii) attribution of helicity like labels  for $\tau$'s in simulated  {\tt HepMC3} format events,
  (iii)  also  $\tau$ polarimetric vectors prepared for user application of New Physics.
\end{abstract}

\section{Introduction}

The {\tt KKMC} \cite{Jadach:1999vf,Jadach:2022mbe}  program for simulation of lepton-pair productions in $e^+e^-$ collisions, {\tt 
TAUOLA} \cite{Jadach:1993hs,Chrzaszcz:2016fte} for $\tau$ lepton decays and {\tt PHOTOS} \cite{Barberio:1993qi,Davidson:2010ew} for 
radiative corrections in decays are long term projects consisting of construction of Monte Carlo generators, but requiring massive 
theoretical effort as well. Projects efforts were extending over more than 40 years of time. Many people were involved in the 
developments, and the programs have been presented in many occasions, in particular at all $\tau$-lepton conferences since the beginning 
of the conference cycle. Now, the necessary and delicate phase of contributing authors generation
change, require the transfer of various skills and expertise. This is major challenge of the forthcoming several years, but it is 
difficult to address.
The presentation of the programs themselves is of limited interest this year, as not much changes were introduced. Instead, 
presentation of associated activities may be of interest. That will be covered in the following sections, each presenting different, 
but correlated with the listed Monte Carlo programs projects.

\section{On feasibility of using KKMC Monte Carlo  for $e^+e^- \to \pi^+\pi^- \gamma$.}
{
An old application for use of {\tt KKMC} for $\pi$-pair production got lost. Only some plots prepared by  S. Jadach survived, see 
\cite{WorkingGrouponRadiativeCorrections:2010bjp} and \cite{Jadach:2005gx}. However interest in the application remains. I was asked 
if something can be done to recover. I found it easier to prepare implementation in a new but distinct manner, more suitable for 
{\tt Belle II} quick implementation. That means   re-weighting of $e^+e^- \to \mu^+\mu^-n \gamma$ (muon mass replaced with $\pi^\pm$ 
mass) into $\pi^+\pi^-$ pairs. We have followed as follows.
   First, only initial state bremsstrahlung was studied, for that purpose  {\tt KKMC}, and several options for QED matrix elements 
were used: the {\tt  CEEX0, CEEX1, CEEX2} and {\tt CEEX2}  with further improvements were compared.
Suitability of the {\tt CEEX2} choice (with or without improvements)  is justified
in  Fig.~\ref{fig:pipi}  where ratio of the $\pi-\pi$-pair invariant mass,
to the one if $\mu\mu$- pair (of $\pi$ mass) is plotted. Modified matrix element was used.
For the hard photon events, angular distribution of the events and events after re-weighting to $\pi$ production (no 
bremsstrahlung) matrix elements was shown.
That  preliminary test, for {\tt KKMC} f77 was encouraging,
but further studies with $\pi$-pair production and comparisons with other
programs (as in  \cite{Jadach:2005gx})  was not completed. Project got stalled.
But now, with  {\tt KKMCee}  C++ version offer a nice project to revive \cite{John:2025ofs}.
This work points to another topic, that means tests and applications for spin effects
in $\tau$ pair production.

\begin{figure}
\centering

\begin{subfigure}{0.32\textwidth}\centering
\resizebox{\linewidth}{!}{%
\setlength{\unitlength}{0.03mm}
\begin{picture}(1600,1500)
\put(300,250){\begin{picture}( 1200,1200)
\put(0,0){\framebox( 1200,1200){ }}
\multiput(  300.00,0)(  300.00,0){   4}{\line(0,1){25}}
\multiput(    0.00,0)(   30.00,0){  41}{\line(0,1){10}}
\multiput(  300.00,1200)(  300.00,0){   4}{\line(0,-1){25}}
\multiput(    0.00,1200)(   30.00,0){  41}{\line(0,-1){10}}
\put( 300,-25){\makebox(0,0)[t]{ $    2.5 $}}
\put( 600,-25){\makebox(0,0)[t]{ $    5.0 $}}
\put( 900,-25){\makebox(0,0)[t]{ $    7.5 $}}
\put(1200,-25){\makebox(0,0)[t]{ $   10.0 $}}
\multiput(0,  320.08)(0,  320.20){   3}{\line(1,0){25}}
\multiput(0,   31.89)(0,   32.02){  37}{\line(1,0){10}}
\multiput(1200,  320.08)(0,  320.20){   3}{\line(-1,0){25}}
\multiput(1200,   31.89)(0,   32.02){  37}{\line(-1,0){10}}
\put(-25, 320){\makebox(0,0)[r]{ $    2 $}}
\put(-25, 640){\makebox(0,0)[r]{ $    4 $}}
\put(-25, 960){\makebox(0,0)[r]{ $    6 $}}
\end{picture}}
\put(300,250){\begin{picture}( 1200,1200)
\thicklines 
\newcommand{\x}[3]{\put(#1,#2){\line(1,0){#3}}}
\newcommand{\y}[3]{\put(#1,#2){\line(0,1){#3}}}
\newcommand{\z}[3]{\put(#1,#2){\line(0,-1){#3}}}
\newcommand{\e}[3]{\put(#1,#2){\line(0,1){#3}}}
\y{   0}{   0}{  19}\x{   0}{  19}{  20}
\y{  20}{  19}{  64}\x{  20}{  83}{  20}
\y{  40}{  83}{ 307}\x{  40}{ 390}{  20}
\y{  60}{ 390}{ 753}\x{  60}{1143}{  20}
\z{  80}{1143}{ 670}\x{  80}{ 473}{  20}
\z{ 100}{ 473}{ 305}\x{ 100}{ 168}{  20}
\z{ 120}{ 168}{  81}\x{ 120}{  87}{  20}
\z{ 140}{  87}{  31}\x{ 140}{  56}{  20}
\z{ 160}{  56}{  16}\x{ 160}{  40}{  20}
\z{ 180}{  40}{  11}\x{ 180}{  29}{  20}
\z{ 200}{  29}{   9}\x{ 200}{  20}{  20}
\z{ 220}{  20}{   8}\x{ 220}{  12}{  20}
\z{ 240}{  12}{   7}\x{ 240}{   5}{  20}
\z{ 260}{   5}{   3}\x{ 260}{   2}{  20}
\z{ 280}{   2}{   1}\x{ 280}{   1}{  20}
\y{ 300}{   1}{   0}\x{ 300}{   1}{  20}
\y{ 320}{   1}{   2}\x{ 320}{   3}{  20}
\y{ 340}{   3}{   3}\x{ 340}{   6}{  20}
\y{ 360}{   6}{   1}\x{ 360}{   7}{  20}
\z{ 380}{   7}{   2}\x{ 380}{   5}{  20}
\z{ 400}{   5}{   2}\x{ 400}{   3}{  20}
\z{ 420}{   3}{   1}\x{ 420}{   2}{  20}
\z{ 440}{   2}{   1}\x{ 440}{   1}{  20}
\y{ 460}{   1}{   0}\x{ 460}{   1}{  20}
\y{ 480}{   1}{   0}\x{ 480}{   1}{  20}
\y{ 500}{   1}{   0}\x{ 500}{   1}{  20}
\y{ 520}{   1}{   0}\x{ 520}{   1}{  20}
\y{ 540}{   1}{   0}\x{ 540}{   1}{  20}
\z{ 560}{   1}{   1}\x{ 560}{   0}{  20}
\y{ 580}{   0}{   0}\x{ 580}{   0}{  20}
\y{ 600}{   0}{   0}\x{ 600}{   0}{  20}
\y{ 620}{   0}{   0}\x{ 620}{   0}{  20}
\y{ 640}{   0}{   0}\x{ 640}{   0}{  20}
\y{ 660}{   0}{   0}\x{ 660}{   0}{  20}
\y{ 680}{   0}{   0}\x{ 680}{   0}{  20}
\y{ 700}{   0}{   0}\x{ 700}{   0}{  20}
\y{ 720}{   0}{   0}\x{ 720}{   0}{  20}
\y{ 740}{   0}{   0}\x{ 740}{   0}{  20}
\y{ 760}{   0}{   0}\x{ 760}{   0}{  20}
\y{ 780}{   0}{   0}\x{ 780}{   0}{  20}
\y{ 800}{   0}{   0}\x{ 800}{   0}{  20}
\y{ 820}{   0}{   0}\x{ 820}{   0}{  20}
\y{ 840}{   0}{   0}\x{ 840}{   0}{  20}
\y{ 860}{   0}{   0}\x{ 860}{   0}{  20}
\y{ 880}{   0}{   0}\x{ 880}{   0}{  20}
\y{ 900}{   0}{   0}\x{ 900}{   0}{  20}
\y{ 920}{   0}{   0}\x{ 920}{   0}{  20}
\y{ 940}{   0}{   0}\x{ 940}{   0}{  20}
\y{ 960}{   0}{   0}\x{ 960}{   0}{  20}
\y{ 980}{   0}{   0}\x{ 980}{   0}{  20}
\y{1000}{   0}{   0}\x{1000}{   0}{  20}
\y{1020}{   0}{   0}\x{1020}{   0}{  20}
\y{1040}{   0}{   0}\x{1040}{   0}{  20}
\y{1060}{   0}{   0}\x{1060}{   0}{  20}
\y{1080}{   0}{   0}\x{1080}{   0}{  20}
\y{1100}{   0}{   0}\x{1100}{   0}{  20}
\y{1120}{   0}{   0}\x{1120}{   0}{  20}
\y{1140}{   0}{   0}\x{1140}{   0}{  20}
\y{1160}{   0}{   0}\x{1160}{   0}{  20}
\y{1180}{   0}{   0}\x{1180}{   0}{  20}
\end{picture}} 
\end{picture} 
}
\caption{ }
\end{subfigure}\hspace{0.3cm}
\begin{subfigure}{0.32\textwidth}\centering
\resizebox{\linewidth}{!}{%
\setlength{\unitlength}{0.03mm}
\begin{picture}(1600,1500)
\put(300,250){\begin{picture}( 1200,1200)
\put(0,0){\framebox( 1200,1200){ }}
\multiput(    0.00,0)(  300.00,0){   5}{\line(0,1){25}}
\multiput(    0.00,0)(   30.00,0){  41}{\line(0,1){10}}
\multiput(    0.00,1200)(  300.00,0){   5}{\line(0,-1){25}}
\multiput(    0.00,1200)(   30.00,0){  41}{\line(0,-1){10}}
\put(   0,-25){\makebox(0,0)[t]{ $  -1 $}}
\put( 300,-25){\makebox(0,0)[t]{ $   -.5 $}}
\put( 600,-25){\makebox(0,0)[t]{$    0.0 $}}
\put( 900,-25){\makebox(0,0)[t]{$    0.5$}}
\put(1200,-25){\makebox(0,0)[t]{ $   1.0 $}}
\multiput(0,  170.78)(0,  354.31){   3}{\line(1,0){25}}
\multiput(0,   29.05)(0,   35.43){  34}{\line(1,0){10}}
\multiput(1200,  170.78)(0,  354.31){   3}{\line(-1,0){25}}
\multiput(1200,   29.05)(0,   35.43){  34}{\line(-1,0){10}}
\put(-25, 171){\makebox(0,0)[r]{ $    2.5\times 10^{   4} $}}
\put(-25, 525){\makebox(0,0)[r]{ $    3.0\times 10^{   4} $}}
\put(-25, 879){\makebox(0,0)[r]{ $    3.5\times 10^{   4} $}}
\end{picture}}
\put(300,250){\begin{picture}( 1200,1200)
\thinlines 
\newcommand{\x}[3]{\put(#1,#2){\line(1,0){#3}}}
\newcommand{\y}[3]{\put(#1,#2){\line(0,1){#3}}}
\newcommand{\z}[3]{\put(#1,#2){\line(0,-1){#3}}}
\newcommand{\e}[3]{\put(#1,#2){\line(0,1){#3}}}
\y{   0}{   0}{1133}\x{   0}{1133}{  20}
\z{  20}{1133}{  60}\x{  20}{1073}{  20}
\z{  40}{1073}{  88}\x{  40}{ 985}{  20}
\z{  60}{ 985}{  50}\x{  60}{ 935}{  20}
\z{  80}{ 935}{  79}\x{  80}{ 856}{  20}
\z{ 100}{ 856}{  81}\x{ 100}{ 775}{  20}
\z{ 120}{ 775}{  39}\x{ 120}{ 736}{  20}
\z{ 140}{ 736}{  81}\x{ 140}{ 655}{  20}
\z{ 160}{ 655}{  50}\x{ 160}{ 605}{  20}
\z{ 180}{ 605}{  70}\x{ 180}{ 535}{  20}
\z{ 200}{ 535}{  47}\x{ 200}{ 488}{  20}
\z{ 220}{ 488}{  54}\x{ 220}{ 434}{  20}
\z{ 240}{ 434}{  23}\x{ 240}{ 411}{  20}
\z{ 260}{ 411}{  39}\x{ 260}{ 372}{  20}
\z{ 280}{ 372}{  68}\x{ 280}{ 304}{  20}
\z{ 300}{ 304}{  32}\x{ 300}{ 272}{  20}
\z{ 320}{ 272}{  23}\x{ 320}{ 249}{  20}
\z{ 340}{ 249}{  63}\x{ 340}{ 186}{  20}
\y{ 360}{ 186}{  11}\x{ 360}{ 197}{  20}
\z{ 380}{ 197}{  62}\x{ 380}{ 135}{  20}
\z{ 400}{ 135}{  13}\x{ 400}{ 122}{  20}
\z{ 420}{ 122}{  17}\x{ 420}{ 105}{  20}
\z{ 440}{ 105}{  32}\x{ 440}{  73}{  20}
\z{ 460}{  73}{  13}\x{ 460}{  60}{  20}
\z{ 480}{  60}{  23}\x{ 480}{  37}{  20}
\z{ 500}{  37}{  22}\x{ 500}{  15}{  20}
\y{ 520}{  15}{  10}\x{ 520}{  25}{  20}
\z{ 540}{  25}{   6}\x{ 540}{  19}{  20}
\z{ 560}{  19}{  19}\x{ 560}{   0}{  20}
\y{ 580}{   0}{  16}\x{ 580}{  16}{  20}
\z{ 600}{  16}{   3}\x{ 600}{  13}{  20}
\z{ 620}{  13}{   4}\x{ 620}{   9}{  20}
\y{ 640}{   9}{   9}\x{ 640}{  18}{  20}
\y{ 660}{  18}{  10}\x{ 660}{  28}{  20}
\z{ 680}{  28}{   5}\x{ 680}{  23}{  20}
\y{ 700}{  23}{  27}\x{ 700}{  50}{  20}
\y{ 720}{  50}{  10}\x{ 720}{  60}{  20}
\y{ 740}{  60}{  18}\x{ 740}{  78}{  20}
\y{ 760}{  78}{  12}\x{ 760}{  90}{  20}
\y{ 780}{  90}{  34}\x{ 780}{ 124}{  20}
\y{ 800}{ 124}{  15}\x{ 800}{ 139}{  20}
\y{ 820}{ 139}{  37}\x{ 820}{ 176}{  20}
\y{ 840}{ 176}{  34}\x{ 840}{ 210}{  20}
\y{ 860}{ 210}{  25}\x{ 860}{ 235}{  20}
\y{ 880}{ 235}{  62}\x{ 880}{ 297}{  20}
\y{ 900}{ 297}{   7}\x{ 900}{ 304}{  20}
\y{ 920}{ 304}{  61}\x{ 920}{ 365}{  20}
\y{ 940}{ 365}{  39}\x{ 940}{ 404}{  20}
\y{ 960}{ 404}{  58}\x{ 960}{ 462}{  20}
\y{ 980}{ 462}{  48}\x{ 980}{ 510}{  20}
\y{1000}{ 510}{  29}\x{1000}{ 539}{  20}
\y{1020}{ 539}{  76}\x{1020}{ 615}{  20}
\y{1040}{ 615}{  63}\x{1040}{ 678}{  20}
\y{1060}{ 678}{  56}\x{1060}{ 734}{  20}
\y{1080}{ 734}{  40}\x{1080}{ 774}{  20}
\y{1100}{ 774}{  53}\x{1100}{ 827}{  20}
\y{1120}{ 827}{  82}\x{1120}{ 909}{  20}
\y{1140}{ 909}{  93}\x{1140}{1002}{  20}
\y{1160}{1002}{  85}\x{1160}{1087}{  20}
\y{1180}{1087}{  56}\x{1180}{1143}{  20}
\end{picture}} 
\end{picture} 
}
\caption{ }
\end{subfigure}
\begin{subfigure}{0.32\textwidth}\centering
\resizebox{\linewidth}{!}{%
\setlength{\unitlength}{0.03mm}
\begin{picture}(1600,1500)
\put(300,250){\begin{picture}( 1200,1200)
\put(0,0){\framebox( 1200,1200){ }}
\multiput(    0.00,0)(  300.00,0){   5}{\line(0,1){25}}
\multiput(    0.00,0)(   30.00,0){  41}{\line(0,1){10}}
\multiput(    0.00,1200)(  300.00,0){   5}{\line(0,-1){25}}
\multiput(    0.00,1200)(   30.00,0){  41}{\line(0,-1){10}}
\put(   0,-25){\makebox(0,0)[t]{ $  -1. $}}
\put( 300,-25){\makebox(0,0)[t]{ $   -.5 $}}
\put( 600,-25){\makebox(0,0)[t]{ $    0.0$}}
\put( 900,-25){\makebox(0,0)[t]{ $    0.5 $}}
\put(1200,-25){\makebox(0,0)[t]{ $   1.0 $}}
\multiput(0,  239.27)(0,  282.21){   4}{\line(1,0){25}}
\multiput(0,   13.50)(0,   28.22){  43}{\line(1,0){10}}
\multiput(1200,  239.27)(0,  282.21){   4}{\line(-1,0){25}}
\multiput(1200,   13.50)(0,   28.22){  43}{\line(-1,0){10}}
\put(-25, 239){\makebox(0,0)[r]{ $    1\times 10^{   4} $}}
\put(-25, 521){\makebox(0,0)[r]{ $    2\times 10^{   4} $}}
\put(-25, 804){\makebox(0,0)[r]{ $    3\times 10^{   4} $}}
\put(-25,1086){\makebox(0,0)[r]{ $    4\times 10^{   4} $}}
\end{picture}}
\put(300,250){\begin{picture}( 1200,1200)
\thicklines 
\newcommand{\x}[3]{\put(#1,#2){\line(1,0){#3}}}
\newcommand{\y}[3]{\put(#1,#2){\line(0,1){#3}}}
\newcommand{\z}[3]{\put(#1,#2){\line(0,-1){#3}}}
\newcommand{\e}[3]{\put(#1,#2){\line(0,1){#3}}}
\y{   0}{   0}{   0}\x{   0}{   0}{  20}
\y{  20}{   0}{  75}\x{  20}{  75}{  20}
\y{  40}{  75}{  72}\x{  40}{ 147}{  20}
\y{  60}{ 147}{  69}\x{  60}{ 216}{  20}
\y{  80}{ 216}{  67}\x{  80}{ 283}{  20}
\y{ 100}{ 283}{  64}\x{ 100}{ 347}{  20}
\y{ 120}{ 347}{  64}\x{ 120}{ 411}{  20}
\y{ 140}{ 411}{  48}\x{ 140}{ 459}{  20}
\y{ 160}{ 459}{  75}\x{ 160}{ 534}{  20}
\y{ 180}{ 534}{  36}\x{ 180}{ 570}{  20}
\y{ 200}{ 570}{  49}\x{ 200}{ 619}{  20}
\y{ 220}{ 619}{  53}\x{ 220}{ 672}{  20}
\y{ 240}{ 672}{  50}\x{ 240}{ 722}{  20}
\y{ 260}{ 722}{  43}\x{ 260}{ 765}{  20}
\y{ 280}{ 765}{  45}\x{ 280}{ 810}{  20}
\y{ 300}{ 810}{  33}\x{ 300}{ 843}{  20}
\y{ 320}{ 843}{  42}\x{ 320}{ 885}{  20}
\y{ 340}{ 885}{  12}\x{ 340}{ 897}{  20}
\y{ 360}{ 897}{  58}\x{ 360}{ 955}{  20}
\y{ 380}{ 955}{  27}\x{ 380}{ 982}{  20}
\y{ 400}{ 982}{  27}\x{ 400}{1009}{  20}
\y{ 420}{1009}{   2}\x{ 420}{1011}{  20}
\y{ 440}{1011}{  22}\x{ 440}{1033}{  20}
\y{ 460}{1033}{  25}\x{ 460}{1058}{  20}
\y{ 480}{1058}{   1}\x{ 480}{1059}{  20}
\y{ 500}{1059}{  39}\x{ 500}{1098}{  20}
\y{ 520}{1098}{  33}\x{ 520}{1131}{  20}
\z{ 540}{1131}{  15}\x{ 540}{1116}{  20}
\z{ 560}{1116}{  14}\x{ 560}{1102}{  20}
\y{ 580}{1102}{   9}\x{ 580}{1111}{  20}
\y{ 600}{1111}{  12}\x{ 600}{1123}{  20}
\y{ 620}{1123}{  20}\x{ 620}{1143}{  20}
\z{ 640}{1143}{  27}\x{ 640}{1116}{  20}
\z{ 660}{1116}{  11}\x{ 660}{1105}{  20}
\z{ 680}{1105}{  20}\x{ 680}{1085}{  20}
\y{ 700}{1085}{   9}\x{ 700}{1094}{  20}
\z{ 720}{1094}{  41}\x{ 720}{1053}{  20}
\z{ 740}{1053}{  18}\x{ 740}{1035}{  20}
\z{ 760}{1035}{   5}\x{ 760}{1030}{  20}
\z{ 780}{1030}{  53}\x{ 780}{ 977}{  20}
\z{ 800}{ 977}{   4}\x{ 800}{ 973}{  20}
\z{ 820}{ 973}{   7}\x{ 820}{ 966}{  20}
\z{ 840}{ 966}{  48}\x{ 840}{ 918}{  20}
\z{ 860}{ 918}{  58}\x{ 860}{ 860}{  20}
\z{ 880}{ 860}{  27}\x{ 880}{ 833}{  20}
\z{ 900}{ 833}{  24}\x{ 900}{ 809}{  20}
\z{ 920}{ 809}{  36}\x{ 920}{ 773}{  20}
\z{ 940}{ 773}{  58}\x{ 940}{ 715}{  20}
\z{ 960}{ 715}{  39}\x{ 960}{ 676}{  20}
\z{ 980}{ 676}{  45}\x{ 980}{ 631}{  20}
\z{1000}{ 631}{  59}\x{1000}{ 572}{  20}
\z{1020}{ 572}{  47}\x{1020}{ 525}{  20}
\z{1040}{ 525}{  51}\x{1040}{ 474}{  20}
\z{1060}{ 474}{  65}\x{1060}{ 409}{  20}
\z{1080}{ 409}{  57}\x{1080}{ 352}{  20}
\z{1100}{ 352}{  70}\x{1100}{ 282}{  20}
\z{1120}{ 282}{  66}\x{1120}{ 216}{  20}
\z{1140}{ 216}{  67}\x{1140}{ 149}{  20}
\z{1160}{ 149}{  75}\x{1160}{  74}{  20}
\z{1180}{  74}{  73}\x{1180}{   1}{  20}
\end{picture}} 
\end{picture} 
}
\caption{ }
\end{subfigure}
\caption{\small (a)Ratio of: reweighted  $\pi^+\pi^-$ invariant mass to generated $\mu^+\mu^-$ invariant mass. (b) The $\cos\theta$ 
 distribution of  $\mu^+\mu^-$ direction. Only  hard photons events selected. {\tt Mustraal} frame used (to be explained later). 
Shape of Born distribution recovered. Hard scattering angle $\theta$ is defined in the muon pair rest frame ({\tt Mustraal} or 
Collins Soper). (c) The same $\cos\theta$ distribution of  $\mu^+\mu^-$, but  reweighted to $\pi^+\pi^-$, as previously, only   
hard photons events selected.
}
\label{fig:pipi}
\end{figure}
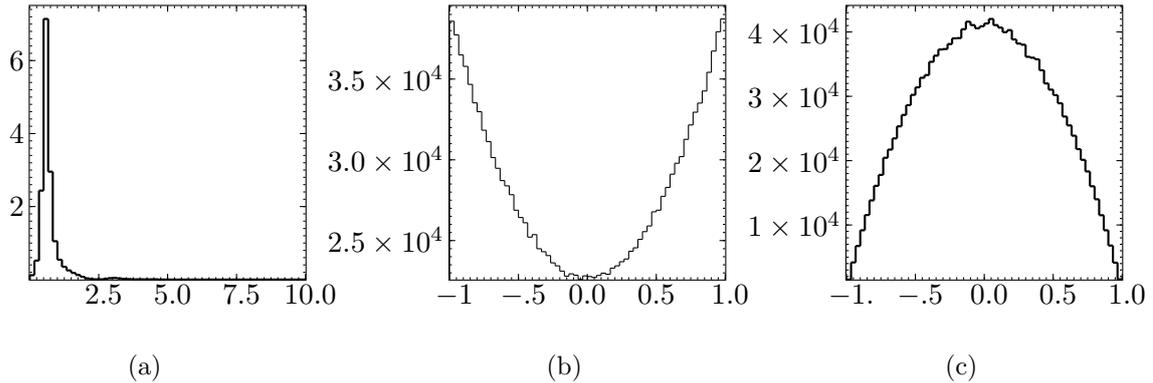

\section{\bf Factorizing distributions.}
The {\tt KKMC} Monte Carlo program represents a refined project based on exclusive exponentiation and on laborious analyses and 
implementation of the first and second order matrix elements. However, for some tests and other applications, a simplified 
picture of factorizing processes into bremsstrahlung part and an effective
Born is enough. We provide only one reference on the topic 
\cite{Richter-Was:2020jlt}; the interested reader is referred to the references therein. One can find there that the studied frames  
found applications outside of QED and are useful for hadronic interactions as well.

Let us return to our topic. In case of the {\tt Mustraal} frame, the $\theta$  distribution
of hard bremsstrahlung events, before reweighting from muons to pions, matched  well the Born  $1+\beta^2 cos^2\theta$ shape, see 
Fig \ref{fig:pipi} (b).
After re-weighting, Born level $\pi\pi$ shape of $\cos\theta$ distribution was recovered too. For the {\tt Collins-Soper} frame the test worked, 
but with slightly lower accuracy.

Now let us address the question what the {\tt Mustraal} frame is, and why it is of interest. Originally, it was devised for ultrarelativistic  
muon pairs with energies around the Z peak. Now we want to  use it for $\tau$'s/$\mu$'s and at relatively low energies\footnote{ We 
leave aside final-state bremsstrahlung (which can be generated with {\tt PHOTOS} MC). Ref.\cite{Nanava:2010xvz} can be 
used as a starting point to study ambiguity. Initial-final-state bremsstrahlung  interference remains to be checked as well. However, we 
can use $\mu$-s generation of KKMC to estimate ambiguity.}. What is the origin of the {\tt Mustraal} frame? Is it only because of some 
lucky properties of matrix elements?
At one time, the {\tt Mustraal} frame \cite{Berends:1983mi}  looked like a technical trick, but it originates from properties of the
Lorentz group representation.
That is why the applicability domain extends, even when a photon is `replaced' by a hadronic jet or  two or more photons/jets. The solution  
works also in the relativistic regime, not only in the ultrarelativistic one. An important observation: ISR  (or FSR) matrix element leads to a sum of two non-coherent
distributions. They can be treated separately. The choice of the {\tt Mustraal} frame exploits this property and is \underline{\it 
useful} not only
for algorithms but for the construction of observables as well.

\subsection{Validating in low energy limit;  {\tt Mustraal} frame when $m_\tau/E >>0$}

The {\tt  Mustraal} frame design was based on matrix elements calculated in the ultrarelativistic limit and for single-photon emission 
only. The solution worked in the ultrarelativistic limit for single and double hadronic jet emission.
That was helpful to evaluate and improve {\tt TauSpinner}, an algorithm useful in {\tt LHC} applications. But what about the lower 
energy regime, like for $\tau$-s at {\tt Belle II} energies? It is an interesting exercise in itself, but also for building intuition. 

{\tt KKMC} has it better, so with it we can check how the factorized Born kinematics works.
In the presence of the ISR, the {\tt Mustraal} frame matches better the theoretical prediction. The plots are made with 10 
million events and the statistical error is not evaluated.

\begin{align}
    \frac{d\sigma^{\langle {Born}\rangle}}{d\cos\theta}
    &= 1 + \cos^{2}\theta
      + \left\langle \frac{m_\tau^{2}}{E^{2}} \right\rangle \sin^{2}\theta
      + \frac{3}{8}\,\langle A_{\mathrm{FB}}\rangle \cos\theta \\
   f_{model1} &= 1 + \cos^{2}\theta_{\mathrm{CS}}
      + F_{1}\sin^{2}\theta_{\mathrm{CS}}
      + F_{2}\cos\theta_{\mathrm{CS}}  \\
    f_{model2} &= 1 + \cos^{2}\theta_{\mathrm{M}}
      + F_{1}\sin^{2}\theta_{\mathrm{M}}
      + F_{2}\cos\theta_{\mathrm{M}}\\
    E_{\tau\tau} &= \frac{\textit{mass of tau pair system}}{2}
\end{align}
  \smallskip
  We effectively sum contributions of the Born from different energies, which is why $\left\langle m_\tau^{2}/E^{2}\right\rangle$ and 
  $\langle A_{\mathrm{FB}}\rangle$ are averaged.
  The $\theta_{\mathrm{CS}}$ and  $\theta_{\mathrm{M}}$ denote  the scattering angle reconstructed  when Collins-Soper
or {\tt Mustraal} frames are used. How well distribution is it reproduced? We use the plots shown in Fig.~\ref{fig:cos} to evaluate.
  The \texttt{<...>} denotes averaging over the photon spectrum generated by \texttt{KKMCee}, and $F_1$, $F_2$ are the results of fit. 

  \begin{figure}[h]
\begin{subfigure}{0.5\textwidth}
\includegraphics[width=\linewidth]{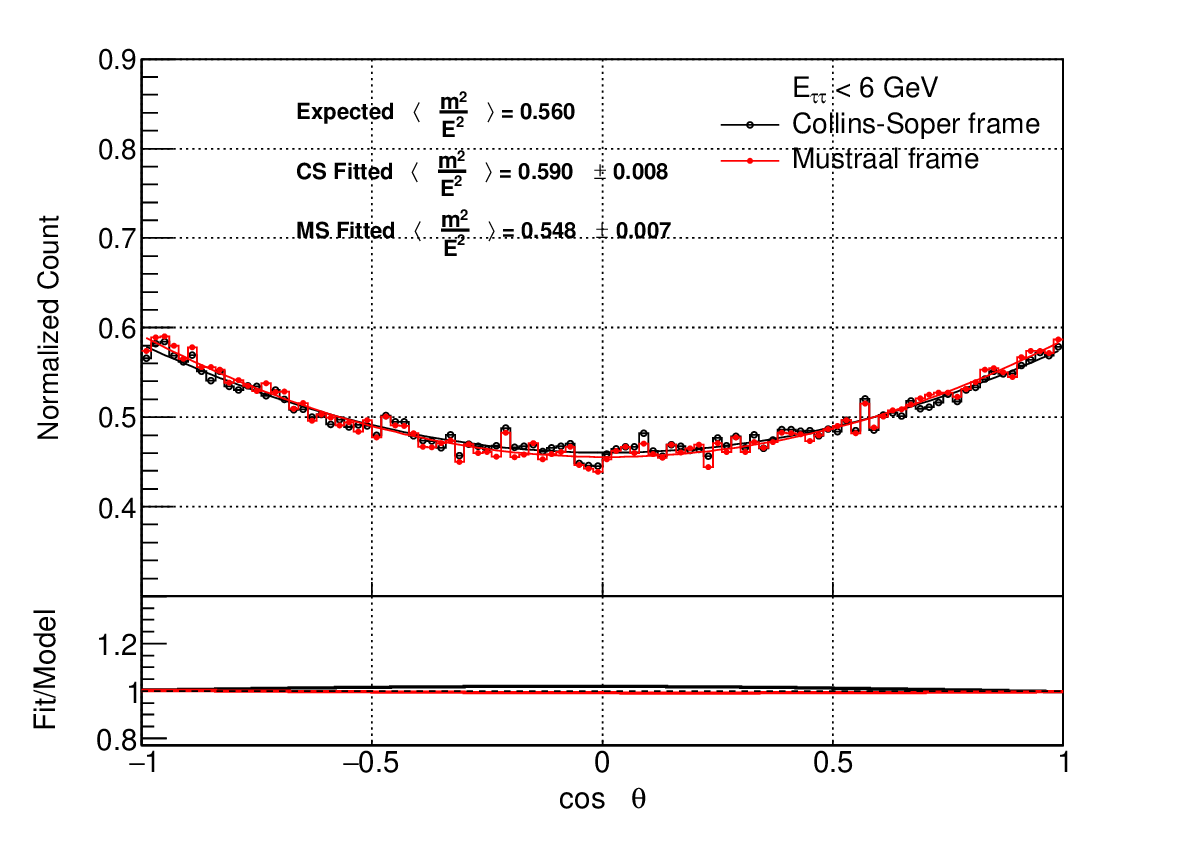} 
\caption{CM energy = 10.58 GeV}
\label{fig:cm_10_58}
\end{subfigure}
\begin{subfigure}{0.5\textwidth}
  \includegraphics[width=\linewidth]{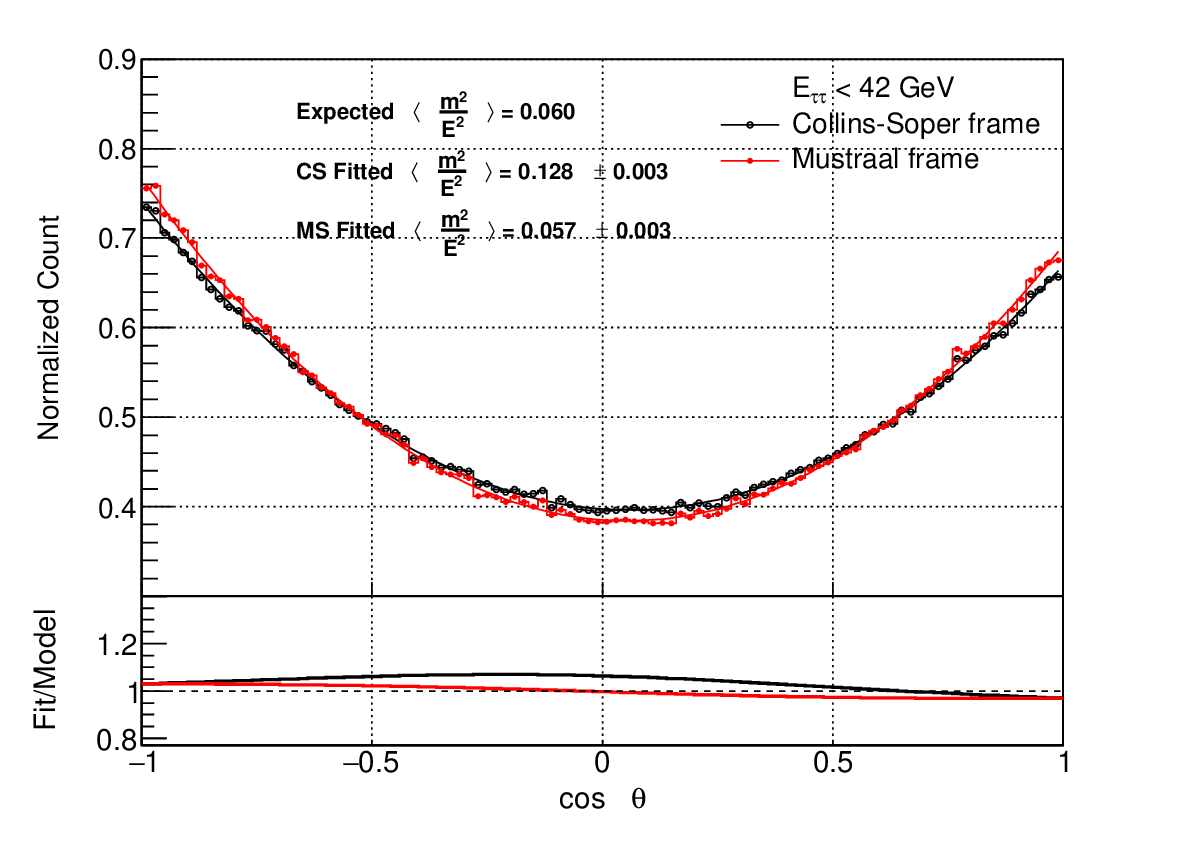} 
\caption{CM energy = 74.06 GeV}
\label{fig:cm_74_06}
\end{subfigure}
\caption{Low energy validation plots: comparison of effective $\cos\theta$ distributions obtained with
  {\tt Collins--Soper} and {\tt Mustraal}  frames with semi-analytical distributions of Born level distribution merged with {\tt 
KKMCee} invariant mass generated spectrum.
}
\label{fig:cos}
\end{figure}

The {\tt Collins--Soper} frame is constructed for LO; no features of radiative corrections are built into the coordinate system 
choice. The {\tt Mustraal} algorithm to reconstruct the event's effective ``Born-like'' reference frame after photon emission (NLO) uses a bit 
more of matrix element properties. We could observe: separation into Born-like variables and photon-emission part works sufficiently well also 
at low energies. One can modify small-scale interactions: i.e. the Born level  differential distribution where e.g, New Physics 
effects can be added. Factorized out bremsstrahlung remains unchanged.
That is  true for the {\tt Collins-Soper} frame, which is conceptually improved leading log(LL). It is also  true for the {\tt Mustraal} 
frame, which takes into account QED single-emission matrix element properties (NLL).

In the following, we will move to the (old) applications, which are now available in
{\tt KKMCee} as well. Note that helicity attribution has already been used in the ALEPH $\frac{g_V}{g_A}$ measurement. Some tests 
and plots describing details (for this type of production/decay factorization) are of interest.  Evaluation of ambiguities and {\tt 
the tools} useful in their evaluation calls for attention as well.

\section{ Applications }
For application purposes we have extended the information stored by {\tt KKMCee} in the event records. Let us have a look at
the following snapshot of {\tt KKMCee}  {\tt HepMC3} event output\footnote{ Appropriate version of {\tt KKMCee}  can currently be
downloaded from:\newline \href{https://holeczek.web.cern.ch/public/KrakowHEPSoft/KKMCee-dev/}{\tt 
https://holeczek.web.cern.ch/public/KrakowHEPSoft/KKMCee-dev/}}.

\includegraphics[width=0.98\linewidth]{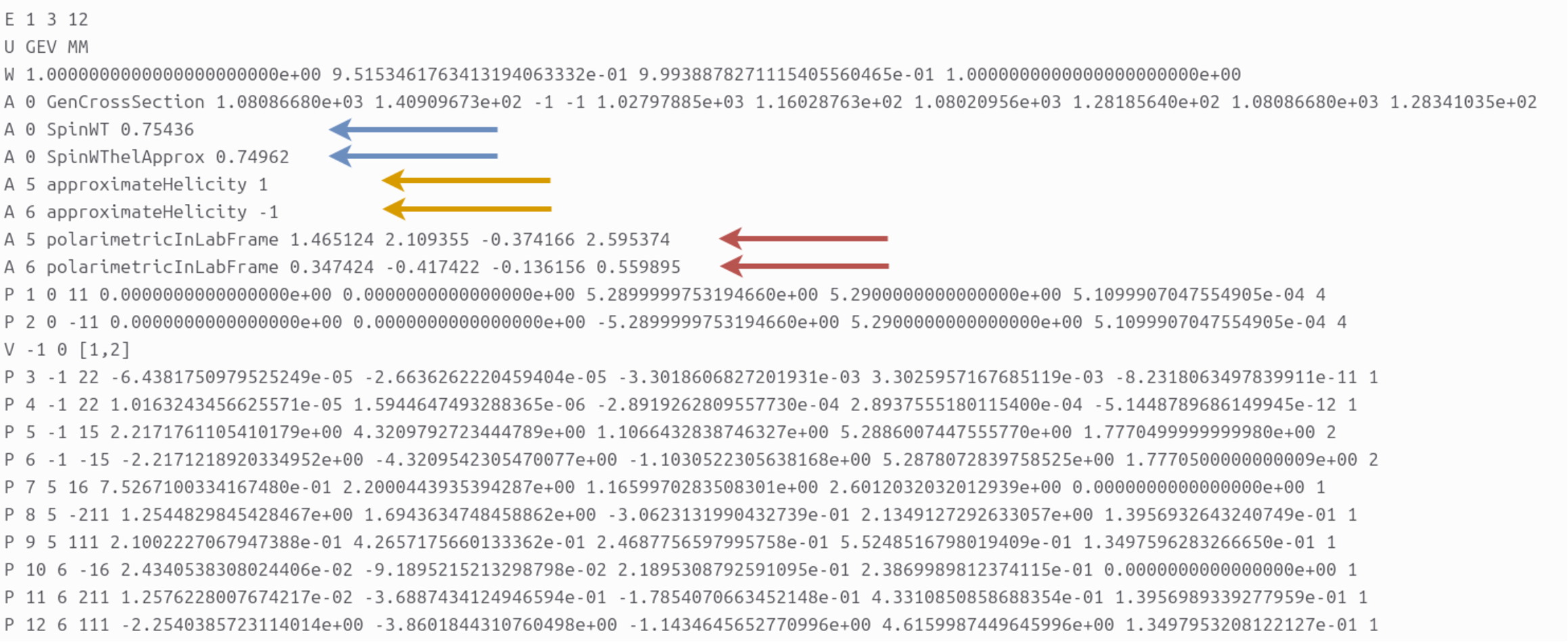}%

Helicity-approximated weights are calculated internally in {\tt KKMC} in parallel with the usual ones used for event construction. For the
helicity approximation, we simply drop transverse spin components. Later, the outer product of two vectors of pure spin states is used: 
for $m/n=+$ :  $(1, 0, 0, 1)$ and for $m/n=-$ : $(1, 0, 0, -1)$, with $p_\tau$ along the z-axis. The  $h^{\tau^+}_{m}$ and $h^{\tau^-}_{n}$ 
and $R_{mn}$ are respectively $\tau$ polarimetric vectors $h$ and $\tau$-pair spin-state $R$ contracted with these pure spin-state 
vectors.

\begin{equation}
  \mathrm{wt}=\sum_{i,j=t,z, {  \slashed{x} \slashed{y}}} h^{\tau^+}_{i}\,R_{ij}\,h^{\tau^-}_{j}
  \;\longrightarrow\;
  \mathrm{wt}=\sum_{m,n=\pm} h^{\tau^+}_{m}\,R_{mn}\,h^{\tau^-}_{n}
\end{equation}

The introduction of these (+/-) vectors enables  the approximated spin weight to be presented as a sum of the helicity/chiral states 
contributions. We project polarimetric vectors along the momenta of $\tau$'s (directions of direct boosts from the lab frame) as used 
internally in {\tt KKMC}. That is not new. This approach was used for the precise measurement of the Weinberg angle from $\tau$ 
polarization by the ALEPH collaboration already. This (+/-) attribution can be done by {\tt KKMC} or externally by the dedicated 
application%
\footnote{The Current version and future updates of the standalone code, working on {\tt HepMC3} events are available from: \href{https://th.ifj.edu.pl/kkmc-demos/index.html}{\tt 
https://th.ifj.edu.pl/kkmc-demos/index.html}. Technical details on how to use it, are given at 
\href{https://th.ifj.edu.pl/kkmc-demos/download.html}{\tt https://th.ifj.edu.pl/kkmc-demos/download.html}}%
. Keep in mind that it is impossible to have helicities for decaying taus without approximation because of quantum entanglement. {\tt 
KKMCee} has its own way of choosing the $z$ direction, also in the case of no bremsstrahlung. Massive smearing of the $z$ direction (adaptation 
to Kleiss--Stirling techniques) with respect to direct direction of boost to lab frame. To overcome this, we had to modify 
{\tt KKMC} to store polarimetric vectors in the lab frame  \texttt{tralo4(-1/-2,\ldots)}. In our external code, the {\tt Mustraal} frame is 
used, $\tau$ momenta are projected along the z-axis. Let us provide one plot, extending validation of the {\tt Mustraal} frame as 
well.

\begin{figure}[h]
\begin{subfigure}{0.5\textwidth}
  \includegraphics[width=\linewidth]{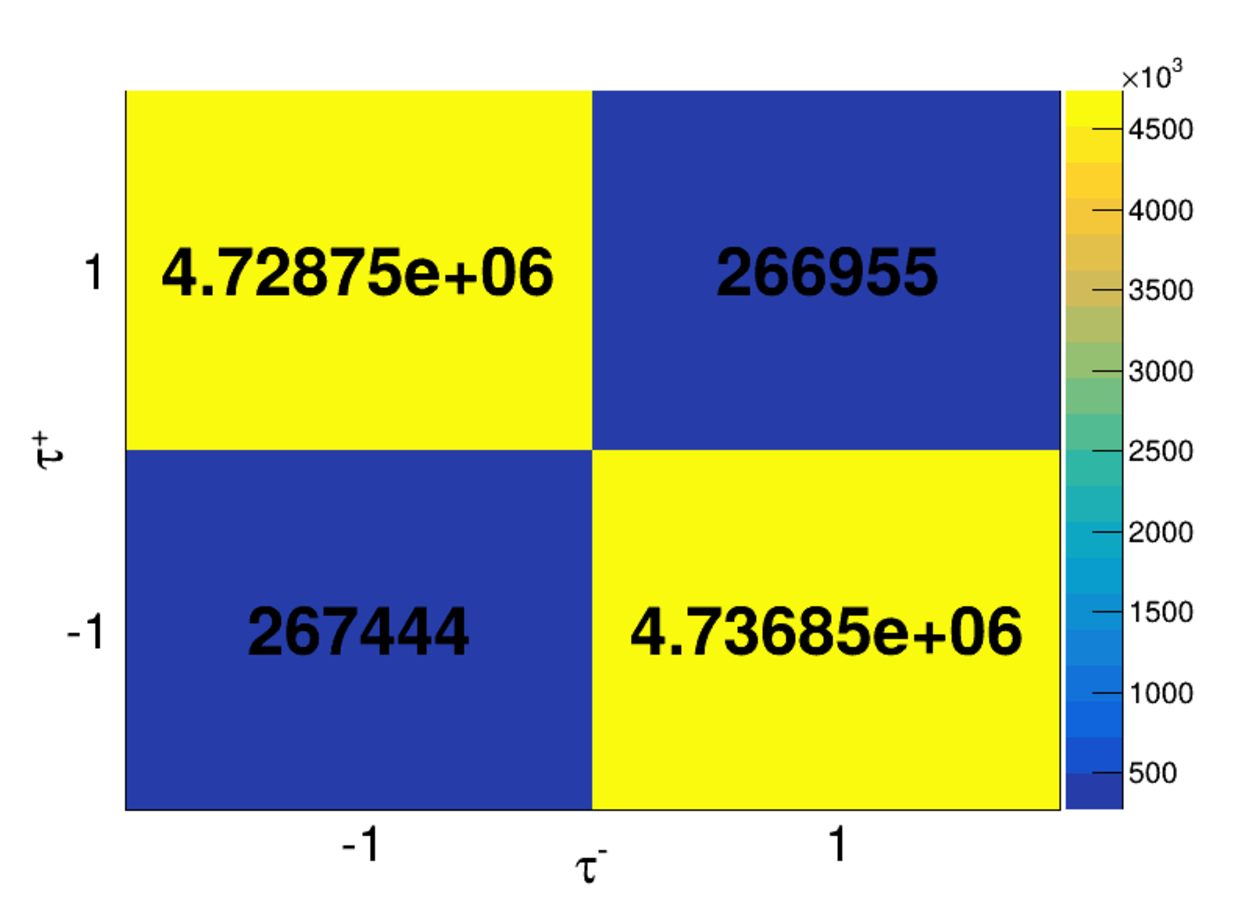} 
\caption{Stand alone code acting on {\tt HepMC3} content}
\label{fig:external_code}
\end{subfigure}
\begin{subfigure}{0.5\textwidth}
  \includegraphics[width=\linewidth]{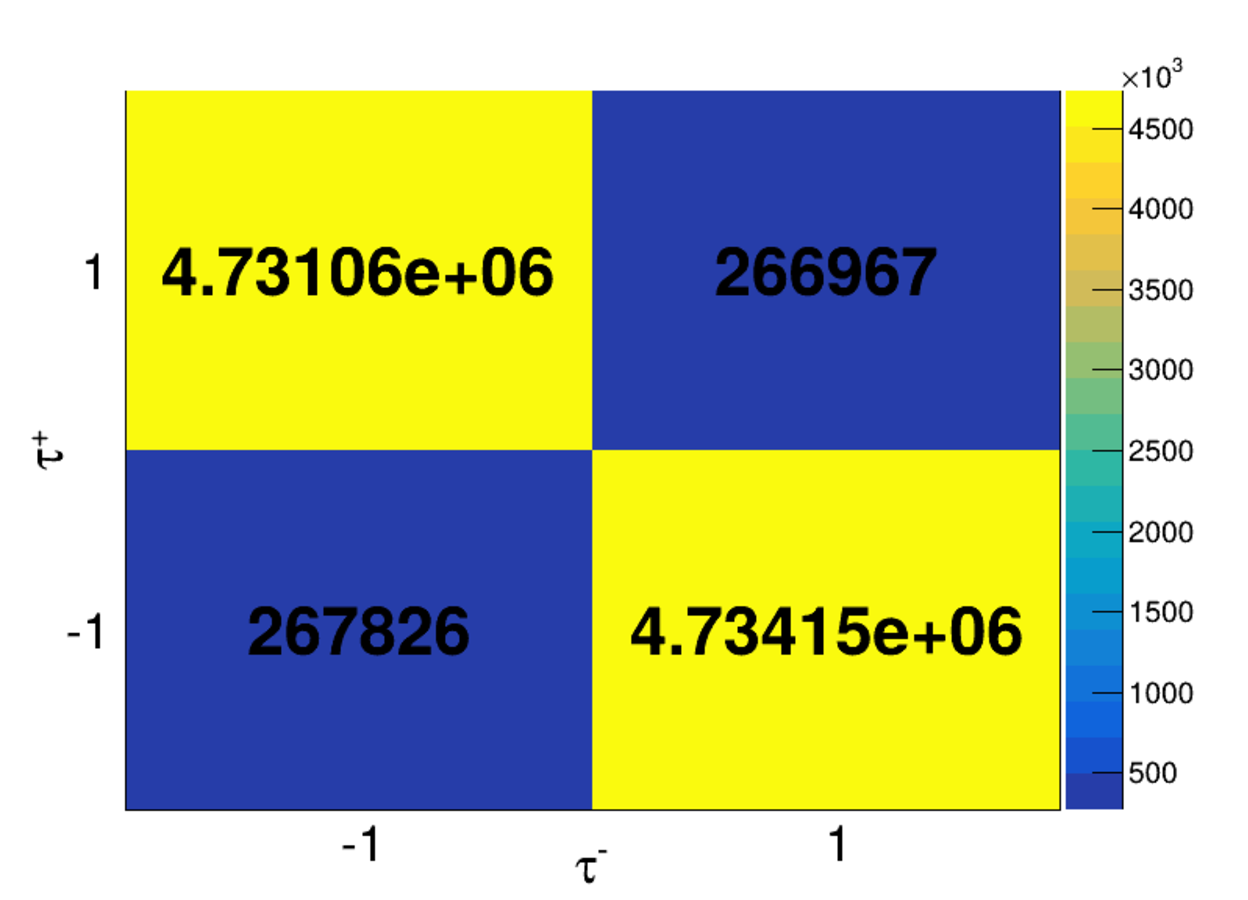} 
\caption{KKMCee $z$ along $\tau$ momentum}
\label{fig:kkmcee}
\end{subfigure}
\caption{The figures are made at CM energy = 10.58 GeV and provide benchmark, both for our stand-alone and {\tt KKMCee} algorithms. The attribited
  helicity-likes are presented.
}
\label{fig:hel_frac}
\end{figure}

Since the external reconstruction reproduces the \textbf{KKMCee helicity fractions} not only  at the \textbf{Z pole} with sub-per-mille 
accuracy, it can now be confidently used to assign helicity states event-by-event, following the same strategy as adopted by the 
ALEPH Collaboration \cite{ALEPH:2001uca}. In particular, the validated external algorithm can serve as an independent helicity 
attribution tool, allowing experimental analyses to extract \(\tau\) polarization and spin-correlation observables directly from 
generated or reconstructed data, without relying on internal {\tt KKMC} routines. The application can attribute helicities for muons as 
well, if such a need arises.

Now let us focus on polarimetric vectors and their application to New Physics searches. They are stored in {\tt HepMC3} format. With 
their help, we could re-do the solution presented in Ref.\cite{Banerjee:2022sgf}, and obtain the same results, see Fig.\ref{fig:acopl}. 
The advantage is that now we can rely on information stored in the event file only. We were not obliged to run the anomalous coupling 
code simultaneously with event generation. That offers flexibility for studies
of distinct New Physics scenarios without the need to re-do 
each time the detector response simulation. The anomalous implementation with event weights can be repeated several times on the same 
production file. This saves user effort and CPU  time as well.

\begin{figure}
  \includegraphics[width=0.98\linewidth]{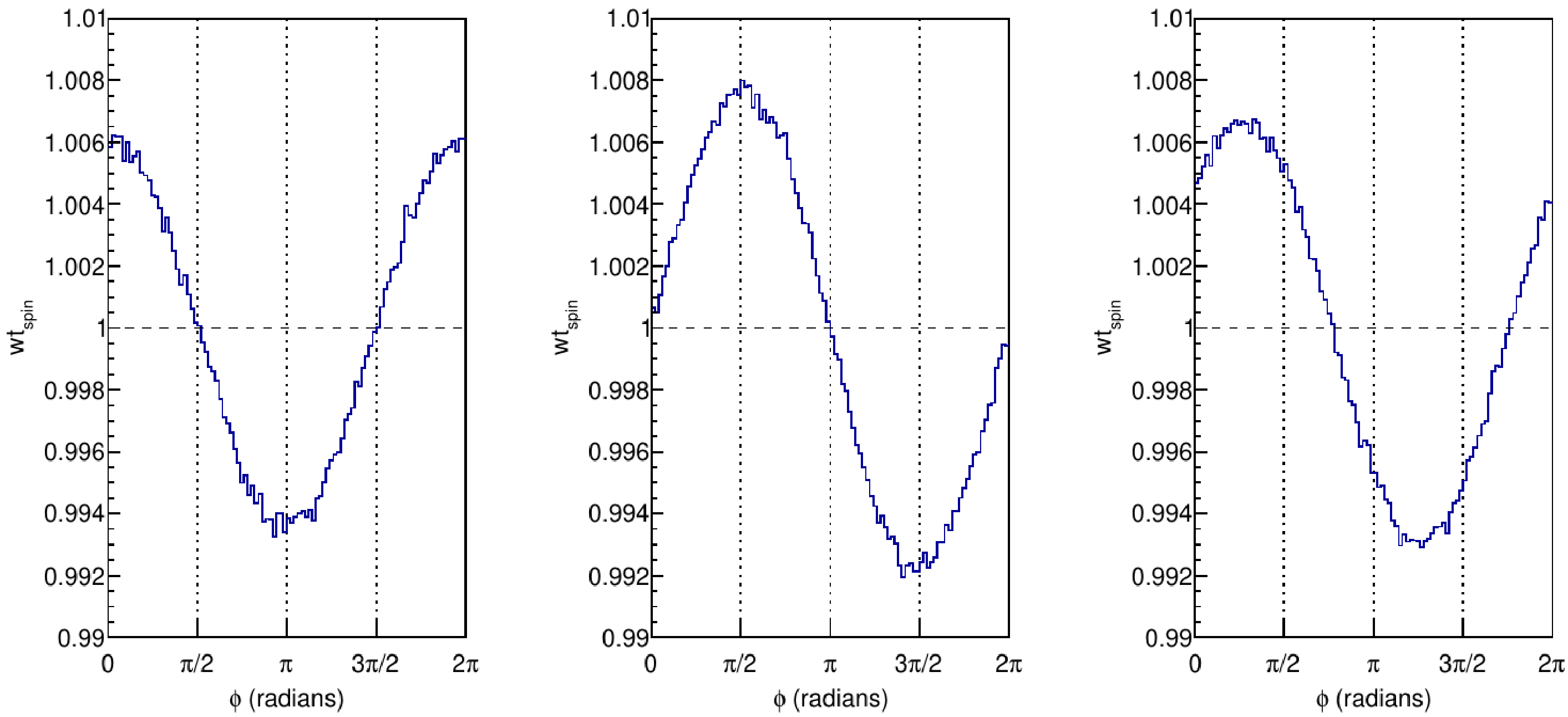}%
 \caption{ Validation result. With stand-alone tool, we have reconstructed  plot of \cite{Banerjee:2022sgf}. \label{fig:acopl}} 
\end{figure}


The standalone code demonstrates consistent behaviour across both longitudinal and transverse spin components, yielding physically reliable 
results over a wide range of energy scales, up to some precision. Let us note that a similar technique, but requiring phase space and 
not only matrix element manipulations, was used in
{\tt PHOTOS} to generate dark photons or light scalars. But this was presented at previous $\tau$ conferences.
\vspace{0.3cm}

There was still another topic covered that was of {\tt Belle II} {\tt TAUOLA} users interest. It was about so-called  $\tau^+$ and $\tau^-$ 
angle $\beta$  distribution asymmetry  in $\tau \to 3\pi \nu$ decays. The  $\beta$ is the angle between the lab frame momentum 
($\vec{n_L}$) and $\vec{q_1} \times \vec{q_2}$, where $q_1$ and $q_2$ are the momenta of the identical hadrons in the hadronic 
rest frame. For identical particles, the momenta were ordered such that $|q_1|<|q_2|$. The users have noted an asymmetry in $\pi, \pi, 
\pi$ decay mode, whereas for the $K, \pi, \pi$ no asymmetry was observed when {\tt CLEO} current is used. As this asymmetry is 
related to $CP$ (equivalently $T$) symmetry breaking, it attracts attention and some results of ref.\cite{Ananya} were recalled. 
Dependencies on distinct models of $\tau$ decays require attention.
   It may be encouraging for future measurements. At present, old {\tt ARGUS} measurements~\cite{ARGUS:1990zee} are available only. These 
decays are of particular interest because they require modelling from low and intermediate-energy QCD, but because of the limited 
number of hadrons, a model-independent representation of hadronic current is possible~\cite{Kuhn:1992nz}. That is particularly 
attractive for studying interactions because non-trivial three-hadron interactions must be taken into account. The initial state is 
well defined by electroweak interactions. Does it  hint at ambiguities in isospin rotations? 

We have observed that the $\beta$ angle asymmetry arises from the interference between the complex parts of the two Breit--Wigner 
amplitudes corresponding to the resonances. The imaginary part reflects the time evolution (or decay dynamics) of the resonant 
particle.
The asymmetry appears  when there are two secondary resonances. In the {\tt CLEO} current for the $K^-, \pi^-, \pi^+$ mode, only 
one resonance is present, so no asymmetry is observed. In contrast, the {\tt TAUOLA} version used in {\tt Belle II}, the latest 
standalone code, and even the 25-year-old version of the
{\tt KORALB} code~\cite{Jadach:1994ps} all consistently show the asymmetry, with no differences among these versions. 
As we could see, the asymmetry may arise or may not arise depending on the details of the current and our physics model 
sophistication. This is an important data point for future improvements which seems to be sometimes taken into account and sometimes 
ignored. Of course, all these effects would disappear in the narrow intermediate-width approximation.
However, if a future, more sophisticated model makes the asymmetry disappear, then it would point (or not point) to  $CT$ violation in 
the hadronic sector. Of course, outcome depends on the comparisons with the data.

\medskip
\medskip
\centerline{\bf Acknowledgements}
\medskip
This work was supported in part by funds from the Polish National Science Centre, Poland,
grant No. 2023/50/A/ST2/00224 and by the U.S. Department of Energy and Research Award No. DE-SC0022350. 

\bibliographystyle{unsrt}
\bibliography{references}

\end{document}